\documentclass[aps,prb,preprint,groupedaddress]{revtex4}

\usepackage{graphicx}
\usepackage{textcomp}
\usepackage{gensymb}
\usepackage{amsmath}
\usepackage{array}

\begin{document}

\title[Mechanical dissipation in MoRe superconducting metal drums]{Mechanical dissipation in MoRe superconducting metal drums}

\author{S. Yanai}
\author{V. Singh}%
\author{M. Yuan}
\author{M. Gely}
\author{S.~J.,~Bosman}
\author{G.~A.,~Steele}
\affiliation{Department of Quantum Nanoscience,
Kavli Institute of Nanoscience, Delft University of Technology, Lorentzweg 1, 2628 CJ Delft, The Netherlands.}

\date{\today}

\begin{abstract}

 We experimentally investigate dissipation in mechanical
 resonators made of a disordered superconducting thin film of
 Molybdenum-Rhenium(MoRe) alloy.  By electrostatically driving the drum
 with a resonant AC voltage, we detect its motion using a
 superconducting microwave cavity.  From the temperature dependence
 of mechanical resonance frequencies and quality factors, we find
 evidence for non-resonant, mechanically active two-level systems
 (TLSs) limiting its quality factor at low temperature.  In addition,
 we observe a strong suppression of mechanical dissipation at large
 mechanical driving amplitudes, suggesting an unconventional
 saturation of the non-resonant TLSs.  These new observations shed
 light on the mechanism of mechanical damping in superconducting
 drums and routes towards understanding dissipation in such
 mechanical systems.

\end{abstract}

\pacs{Valid PACS appear here}
\maketitle

Nanoelectromechanical systems have   evolved into an important platform in modern information technology.  They are extensively used for
applications in sensing, filtering and
timing\cite{ekinci_nanoelectromechanical_2005}. One remarkable example
is cavity
opto/electro-mechanics\cite{aspelmeyer_cavity_2014,metcalfe_applications_2014}. The
demonstrations of the quantum ground state of mechanical
resonators have opened new applications of NEMS devices in quantum
information technology\cite{teufel_sideband_2011,chan_laser_2011}. To
this end, the approach of cavity optomechanics which uses the
interaction between light and mechanical motion, has been very
successful and enabled the applications of NEMS towards the
near-quantum limited frequency
conversion\cite{andrews_bidirectional_2014,bochmann_nanomechanical_2013},
temporal and spectrum shaping of
signals\cite{andrews_quantum-enabled_2015}, and a nearly quantum limited
frequency-mixer \cite{lecocq_mechanically_2015}. 

A successful implementation of an optomechanical system is realized by
coupling a superconducting drumhead resonator to a microwave
cavity. For quantum-limited performance of such a coupled system, both
the drumhead resonator and the superconducting cavity
should have low dissipation rates.  In recent years, superconducting metal drums\cite{teufel_circuit_2011}
have emerged as a popular platform for microwave optomechanics. While
such drums can exhibit very low dissipation, there is also a large
spread in reported mechanical Q-factors \cite{teufel_circuit_2011,suh_thermally_2012,wollman_quantum_2015,pirkkalainen_squeezing_2015} and not many reports studying the
dissipation mechanisms in such devices.

Here, we explore mechanical dissipation mechanisms in such
superconducting drum resonators as a function of temperature and
driving amplitude.  The variation of dissipation rate and resonant
frequency with temperature suggest that mechanically active two-level systems
(TLSs)\cite{anderson_anomalous_1972,phillips_two-level_1987} play an
important role, setting the dissipation in these disordered
superconductors akin to acoustic studies performed earlier on
superconducting glasses \cite{raychaudhuri_low_1984}. By varying the
acoustic excitation strength, we further observe an amplitude
dependent damping rate supporting the role of TLSs, similar to the
observations made in superconducting microwave resonators in response
to the electromagnetic
field\cite{gao_experimental_2008,pappas_two_2011} with electrical TLSs, but with an unconventional saturation of the non-resonant mechanical TLSs by the mechanical drive.


The drums studied in this letter were made using films of a superconducting
alloy of Molybdenum and Rhenium (MoRe 60-40).  The
compatibility of MoRe with HF, oxygen plasma, and an elastic modulus
of $\approx$1~GPa makes it an attractive candidate for making hybrid
electromechanical devices\cite{leonhardt_investigation_1999}.  The
electrical properties of MoRe are well studied establishing its
disordered nature with a residual resistance ratio of approximately
unity and a superconducting transition temperature of
9.2~K\cite{lerner_magnetic_1967,seleznev_deposition_2008,sundar_electrical_2013,aziz_molybdenum-rhenium_2014}.
The electrical dissipation of such films in microwave frequency domain
has been characterized in earlier studies\cite{yasaitis_microwave_1975}, as well as recent reports in
coplanar waveguides
\cite{singh_molybdenum-rhenium_2014}.

Fig.~1(a) shows an optical microscope image of our complete
optomechanical device. It consists of a superconducting drumhead
resonator and a high-impedance microwave cavity both made of MoRe.
The mechanically compliant drumhead resonator is galvanically shorted to the
high-impedance microwave cavity, enabling electrostatic actuation of
its motion. The microwave cavity is coupled through the drumhead to the feedline, such that its response can accessed  in a reflection measurement.  Fig.~1(b) shows a scanning electron
microscope of the drumhead resonator.  We apply microwave signals to
the cavity via mechanically compliant capacitor.  Detection of the
motion of the drum occurs through its modulation of the cavity frequency,
$\omega_{c}$, as well as the external cavity decay rate, $\kappa_{e}$
as schematically shown in Fig.~1(c).

To actuate the drumhead resonator, we apply a DC signal $V_{dc} $ and
a small RF signal $V_{ac}$ near the mechanical resonance frequency
$\omega_m$ simultaneously to the input port. Due to capacitive
attraction, this signal exerts a force $C_g^{'}V_{dc}V_{ac}$ on the
drumhead resonator, where $C_g^{'} = dC_g / dx$ is the derivative of
the capacitance between the resonator and the feedline with respect to
distance.  In order to read out the mechanical motion, we drive the
system with a microwave tone at the cavity resonance frequency
$\omega_{c}$. Due to electro-mechanical coupling, mechanical motion
modulates the intra-cavity power, creating sideband signals in the
reflected signal. The sideband signals are amplified and then mixed
down with a local oscillator tone at the cavity resonance
frequency. The signal is further amplified and sent to a spectrum
analyzer.  Using the mechanical resonator as the coupling capacitor to
the cavity enables both direct electrostatic actuation of the motion
and tuning of the mechanical resonance frequency using voltages applied to the feedline.

The fabricated samples are placed in a radiation-tight box and cooled
down to 20~mK in a dilution refrigerator with sufficient attenuation
at each temperature stages to thermalize the microwave signals (see Supplementary Materials
for measurement chain schematic (SM)). We first begin by characterizing the
microwave cavity. The microwave cavity has has a resonance frequency
of $\omega_c=2\pi\times$6.30~GHz, external coupling rate
$\kappa_e=2\pi\times$31.0~MHz, and internal dissipation rate of
$\kappa_i=2\pi\times$25.8~MHz (see SM for
detailed measurements).  The red curve in Fig.~2(a) shows the measured
mechanical response of the resonator along with a skewed-Lorentzian fit ( light-blue line). The slight asymmetry in
the measured homodyne signal arises from the finite electrical
isolation and is discussed in the supplementary material.  From the
fit, we find a mechanical resonance frequency of $\omega_m$ =
7.2885~MHz with a quality-factor $Q_m$ of $50 \times 10^3$ at
$V_{dc}=10$~V. Fig.~2(b) shows a colorscale plot of mechanical
response as a function of frequency of the RF signal and DC voltage
applied to the feedline using a bias tee.  The sharp change in color
reflects the mechanical resonance frequency. As the DC voltage is
tuned away from zero, the mechanical resonance frequency decreases
quadratically, showing the well-studied capacitive softening
effect\cite{kozinsky_tuning_2006}. The mechanical frequency is pulled by
200~kHz for gate voltages of 20~V. The mechanical signal is no longer
visible around zero gate voltage due to the vanishing electrostatic
force.


In Fig.\ 3, we investigate the temperature dependence of the
mechanical response from 23~mK to 1.5~K. We measured the mechanical
resonance frequencies and the quality-factors at different temperatures and
at different applied dc voltages, $V_g=$~7, 14, and 28~V.  Fig~3(a)
shows the normalized shift in the resonance frequency for various
temperature points. As the temperature is increased the resonance
frequency increases logarithmically up to a cross-over temperature of
$\approx$~900~mK. At higher temperatures, we see a slight drop 
in the resonance frequency. Fig.~3(b) shows the quality-factor $Q_m$
change as a function of temperature. As the temperature is increased
from 23 mK, $Q_m$ shows a sharp decrease for all gate
voltages. Above the approximate cross-over temperature observed in the
mechanical frequency, $Q_m$ stops decreasing and saturates at a value
around 10,000.


The logarithmic increase in the frequency shift suggests the presence
of two-level systems\cite{raychaudhuri_low_1984,hoehne_damping_2010,venkatesan_dissipation_2010,imboden_dissipation_2014}.
TLSs can have a very broad spectral
distribution\cite{esquinazi_tunneling_1998}. At temperatures
$k_BT >> \hbar\omega_m$, the resonant TLSs are expected to be
saturated, and not able to contribute to mechanical
dissipation.  However, coupling of the mechanical motion to higher
energy, off-resonant TLSs can still have a significant contribution to
the frequency shift.  Comparing results at three different voltages,
the normalized shifts are independent of mechanical resonant frequency
below the cross over temperature.  Such a temperature dependence can also
be interpreted in the context of a TLSs model: at high temperatures,
part of the mechanical restoring force arises from the dispersive
shift of the thermal population of the high frequency TLSs.  Beyond
the cross-over temperature, these TLSs decouple from the mechanics due
to either changes of their thermal populations or the relaxation rate. As the TLSs are decoupled
the mechanical spring constant reduces, giving a lower mechanical
frequency.  For an off-resonant dispersive interaction, the normalized
frequency shift is expected to scale as $\delta f/f_0 = C_s
\log(T/T_0)$, where $C_s$ is a constant proportional to the filling
factor and TLSs loss tangent\cite{esquinazi_tunneling_1998}. For data
shown in Fig 3(a), we find $C_s\approx4\times10^4$, similar to
previously reported values for mechanical TLSs in disordered superconducting
films\cite{raychaudhuri_low_1984,esquinazi_acoustic_1986}.

To compare the behavior of dissipation with the frequency shift, we
plot $Q_m^{-1}$ in the subpanel of Fig.~3(b). In lower temperature
ranges, we observe an increase in the mechanical dissipation rate
with temperature, which slows down as the temperature approaches
$\approx$~700~mK. As discussed above, the interaction with resonant TLSs can be
neglected due the low frequency of the drum ($k_BT >> \hbar\omega_m$).
Non-resonant TLSs, however, can also result in dissipation due to the
lag between the dispersive shift of their energies due to the
mechanical coupling and their equilibration time with the bath.  The
contribution of the off-resonant interaction to the damping scale as
$Q_m^{-1}=C_s\frac{\Gamma(T)}{\omega_m}$ for $\omega_m>\Gamma(T)$,
where $\Gamma$ is the TLSs relaxation
rate\cite{esquinazi_tunneling_1998}. The behavior observed here of
mechanical dissipation suggest that TLSs relaxation rate increases
linearly with the temperature.

In Fig.~4, we explore saturation effects
of the TLSs in these drums by applying a large mechanical driving
force.  To increase the acoustic excitation strength, we varied the $AC$
driving voltage for mechanical actuation. Fig.~4(a) shows mechanical
responsivity (amplitude/ driving force) at different driving voltages
in the linear and non-linear limits. Within the linear limit we see
an increase in the responsivity as the drive signal is increased,
suggesting an increase in the mechanical quality factor $Q_m$. The
extracted quality factor is shown in Fig.~4(b). In the nonlinear limit (orange trace), we
can also qualitatively deduce increase in $Q_m$ from the increase in
amplitude of the responsivity.  Similar negative nonlinear damping characteristics were also observed in a similar second device.

While a decrease in the mechanical damping shown in figure 4 is
similar to the case of the saturation of resonant TLSs in
superconducting microwave cavities, such saturation effects are not
typically observed when the interaction with the TLSs is non-resonant,
as the non-resonant drive is not able to excite the TLSs directly. The
observation presented here of decreased damping at large mechanical
excitation, also recently reported for the case of graphene resonators
\cite{singh_negative_2015}, suggests that a strongly non-equilibrium
population of high frequency TLSs is induced by the low frequency driving
forces, for example, by either strong higher-order excitation
processes, or by a decoupling of the non-resonant TLSs from their
bath.

In conclusion, we have studied dissipation in the mechanical drumhead
resonators made of superconducting alloy of MoRe.  The temperature
dependence of the dissipation and resonant frequency strongly suggest
the presence of mechanically active TLSs in these disordered
superconducting thin film mechanical resonators. At low temperatures
the main contribution to dissipation and frequency shift stems from
the dispersive interaction with TLSs, with slow relaxation rates
$<$~7~MHz. We further explored at the mechanical dissipation while
varying the strength of acoustic field and observe an amplitude
dependent damping, suggesting a non-equilibrium population of
non-resonant TLSs induced by mechanical drive.

{\bf Supplementary material} See supplementary material for device fabrication steps, cavity characterization, measurement setup and estimation of the mechanical amplitude.

{\bf Acknowledgments} The work was supported by the Dutch Science Foundation (NWO/FOM).

\newpage

\begin{figure}
\includegraphics[width=120mm]{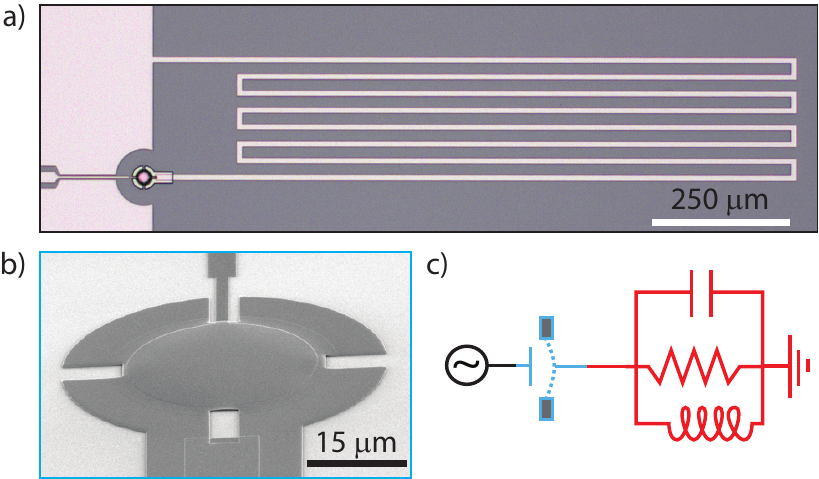}
\caption{Microwave cavity readout of a superconducting drum with
  electrostatic driving.  a) Optical microscope image of the device. A
  drumhead mechanical oscillator is capacitively coupled to the
  microwave input port of a high impedance microwave cavity on a
  sapphire substrate.  b) Scanning electron microscope image of MoRe
  drumhead resonator.  The drum is 30~ $\mu$m in diameter and is
  suspended approximately 290~nm above the gate bottom electrode.  c)
  Device schematic diagram: the mechanical drum is capacitively
  coupled to the microwave input port. Motion of the drum modulates
  both the resonance frequency $\omega_{c}$ and the external coupling
  rate $\kappa_{e}$ of the cavity. }\label{fig 1}
\end{figure}

\begin{figure}
\includegraphics[width=120mm]{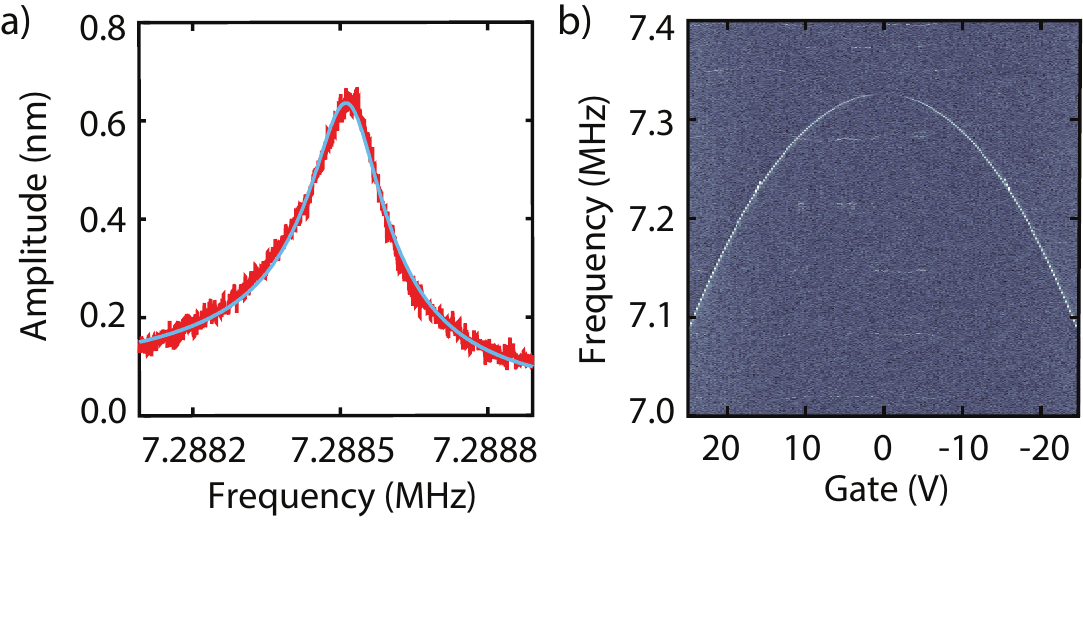}
\caption{Characterization of the mechanical response of the drumhead resonator.   a) Mechanical response of MoRe drumhead resonator at 10~V of applied voltage (red curve) along with the fitted curve (light blue), yielding a quality-factor of 50248 and resonant frequency of 7.2885~MHz. b) Colorscale plot of the measured response with drive frequency and applied voltage. Mechanical response can be tuned over 200~kHz with $\pm$28~V of applied voltage.} 
\end{figure}

\begin{figure}
\includegraphics[width=80mm]{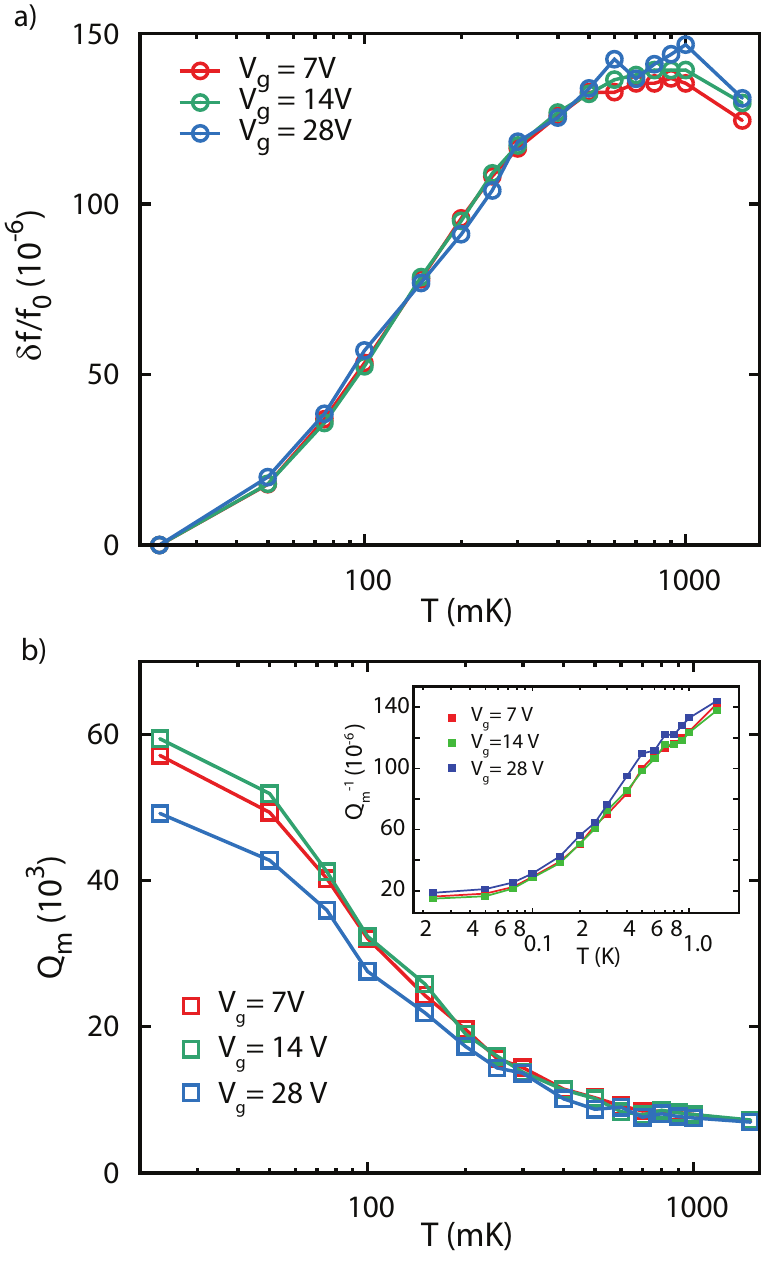}
\caption{Temperature dependence of a) Normalized relative frequency shift $\delta f= (f_0(T)-f_0(23~\text{mK}))/f_0(23~\text{mK})$ and b) the mechanical quality factor. The mechanical quality-factor is determined from fitting to a Lorentzian function. Measurements are taken at three different voltages 7, 14, 28~V. The inset shows the plot of inverse quality-factor $Q_m^{-1}$.}
\end{figure}

\begin{figure}
\includegraphics[width=80mm]{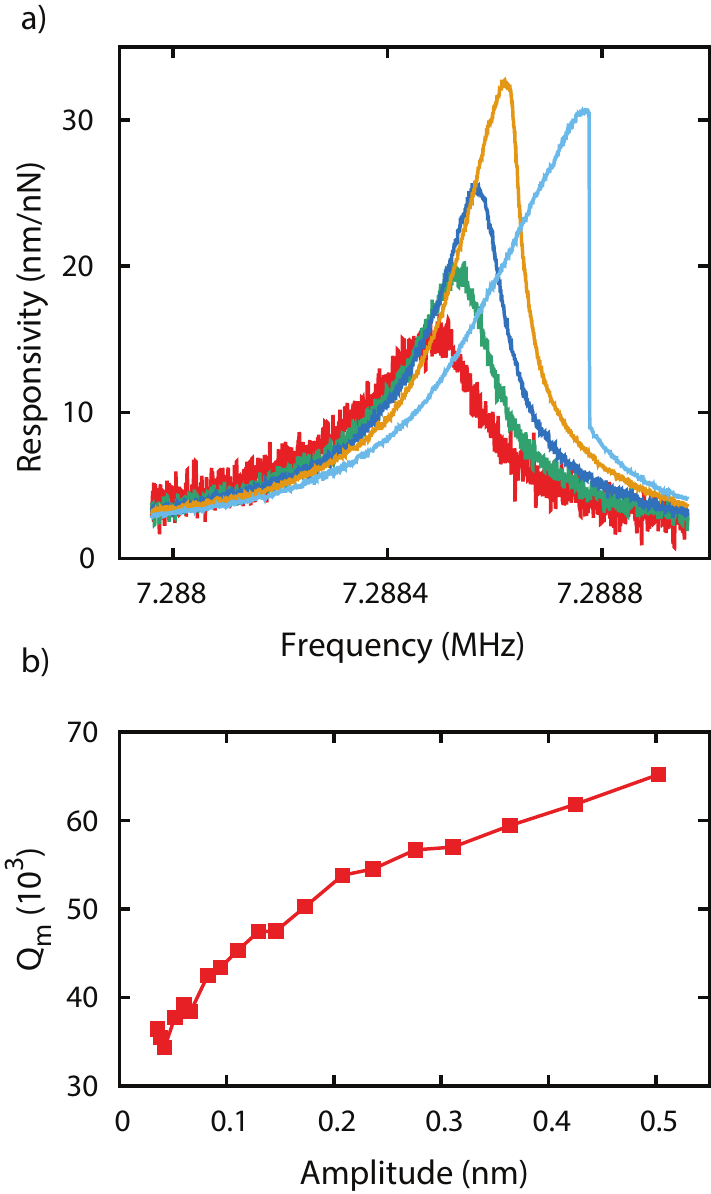}
\caption{Negative nonlinear damping of a superconducting metal drum.  Mechanical responsivity $(x_0 / F_{0})$ of the drum for different driving forces (Red - light blue: 2.9, 5.8, 11.2, 29.3, 82.8~ pN).  As the driving force is increased (red-dark blue), the responsivity of the drum on resonance increases, indicating a increase in the mechanical quality-factor. As the drum is driven into the nonlinear regime (yellow), the $Q _{m}$ continues to increase, and at higher powers, the $Q _{m}$ in the nonlinear regime begins to drop, as can be seen by the decreased responsivity of the light blue curve. In the linear regime, $Q _{m}$ is extracted by fitting the curves with a Lorentzian curve with a Fano correction. b) Mechanical quality-factor as a function of mechanical amplitudes. Actuation force is varied with different RF power on a signal generator for driving mechanics.}
\end{figure}


\begin{thebibliography}{10}

\bibitem{ekinci_nanoelectromechanical_2005}
Ekinci, K.~L. and Roukes, M.~L.
\newblock {\em Review of Scientific Instruments}{ \bf 76}(6), 061101 June
  (2005).

\bibitem{aspelmeyer_cavity_2014}
Aspelmeyer, M., Kippenberg, T.~J., and Marquardt, F.
\newblock {\em Reviews of Modern Physics}{ \bf 86}(4), 1391--1452 December
  (2014).

\bibitem{metcalfe_applications_2014}
Metcalfe, M.
\newblock {\em Applied Physics Reviews}{ \bf 1}(3), 031105 September  (2014).

\bibitem{teufel_sideband_2011}
Teufel, J.~D., Donner, T., Li, D., Harlow, J.~W., Allman, M.~S., Cicak, K.,
  Sirois, A.~J., Whittaker, J.~D., Lehnert, K.~W., and Simmonds, R.~W.
\newblock {\em Nature}{ \bf 475}(7356), 359--363 July  (2011).

\bibitem{chan_laser_2011}
Chan, J., Alegre, T. P.~M., Safavi-Naeini, A.~H., Hill, J.~T., Krause, A.,
  Gröblacher, S., Aspelmeyer, M., and Painter, O.
\newblock {\em Nature}{ \bf 478}(7367), 89--92 October  (2011).

\bibitem{andrews_bidirectional_2014}
Andrews, R.~W., Peterson, R.~W., Purdy, T.~P., Cicak, K., Simmonds, R.~W.,
  Regal, C.~A., and Lehnert, K.~W.
\newblock {\em Nature Physics}{ \bf 10}(4), 321--326 April  (2014).

\bibitem{bochmann_nanomechanical_2013}
Bochmann, J., Vainsencher, A., Awschalom, D.~D., and Cleland, A.~N.
\newblock {\em Nature Physics}{ \bf 9}(11), 712--716 November  (2013).

\bibitem{andrews_quantum-enabled_2015}
Andrews, R.~W., Reed, A.~P., Cicak, K., Teufel, J.~D., and Lehnert, K.~W.
\newblock {\em Nature Communications}{ \bf 6}, 10021 November  (2015).

\bibitem{lecocq_mechanically_2015}
Lecocq, F., Clark, J.~B., Simmonds, R.~W., Aumentado, J., and Teufel, J.~D.
\newblock {\em arXiv:1512.00078 [quant-ph]}{ \bf } November  (2015).
\newblock arXiv: 1512.00078.

\bibitem{teufel_circuit_2011}
Teufel, J.~D., Li, D., Allman, M.~S., Cicak, K., Sirois, A.~J., Whittaker,
  J.~D., and Simmonds, R.~W.
\newblock {\em Nature}{ \bf 471}(7337), 204--208 March  (2011).

\bibitem{suh_thermally_2012}
Suh, J., Shaw, M.~D., LeDuc, H.~G., Weinstein, A.~J., and Schwab, K.~C.
\newblock {\em Nano Letters}{ \bf 12}(12), 6260--6265 December  (2012).

\bibitem{wollman_quantum_2015}
Wollman, E.~E., Lei, C.~U., Weinstein, A.~J., Suh, J., Kronwald, A., Marquardt,
  F., Clerk, A.~A., and Schwab, K.~C.
\newblock {\em Science}{ \bf 349}(6251), 952--955 August  (2015).

\bibitem{pirkkalainen_squeezing_2015}
Pirkkalainen, J.-M., Damskagg, E., Brandt, M., Massel, F., and Sillanpaa, M.
\newblock {\em Physical Review Letters}{ \bf 115}(24), 243601 December  (2015).

\bibitem{anderson_anomalous_1972}
Anderson, P.~w., Halperin, B.~I., and Varma, c.~M.
\newblock {\em Philosophical Magazine}{ \bf 25}(1), 1--9 January  (1972).

\bibitem{phillips_two-level_1987}
Phillips, W.~A.
\newblock {\em Reports on Progress in Physics}{ \bf 50}(12), 1657 December
  (1987).

\bibitem{raychaudhuri_low_1984}
Raychaudhuri, A.~K. and Hunklinger, S.
\newblock {\em Zeitschrift für Physik B Condensed Matter}{ \bf 57}(2),
  113--125 October  (1984).

\bibitem{gao_experimental_2008}
Gao, J., Daal, M., Vayonakis, A., Kumar, S., Zmuidzinas, J., Sadoulet, B.,
  Mazin, B.~A., Day, P.~K., and Leduc, H.~G.
\newblock {\em Applied Physics Letters}{ \bf 92}(15), 152505 April  (2008).

\bibitem{pappas_two_2011}
Pappas, D.~P., Vissers, M.~R., Wisbey, D.~S., Kline, J.~S., and Gao, J.
\newblock {\em IEEE Transactions on Applied Superconductivity}{ \bf 21}(3),
  871--874 June  (2011).

\bibitem{leonhardt_investigation_1999}
Leonhardt, T., Carlén, J.-C., Buck, M., Brinkman, C.~R., Ren, W., and Stevens,
  C.~O.
\newblock In {\em {AIP} {Conference} {Proceedings}}, volume 458,  685--690. AIP
  Publishing,  January  (1999).

\bibitem{lerner_magnetic_1967}
Lerner, E., Daunt, J.~G., and Maxwell, E.
\newblock {\em Physical Review}{ \bf 153}(2), 487--492 January  (1967).

\bibitem{seleznev_deposition_2008}
Seleznev, V.~A., Tarkhov, M.~A., Voronov, B.~M., Milostnaya, I.~I., Lyakhno,
  V.~Y., Garbuz, A.~S., Mikhailov, M.~Y., Zhigalina, O.~M., and Gol’tsman,
  G.~N.
\newblock {\em Superconductor Science and Technology}{ \bf 21}(11), 115006
  November  (2008).

\bibitem{sundar_electrical_2013}
Sundar, S., Sharath~Chandra, L.~S., Sharma, V.~K., Chattopadhyay, M.~K., and
  Roy, S.~B.
\newblock {\em AIP Conference Proceedings}{ \bf 1512}(1), 1092--1093 February
  (2013).

\bibitem{aziz_molybdenum-rhenium_2014}
Aziz, M., Hudson, D.~C., and Russo, S.
\newblock {\em Applied Physics Letters}{ \bf 104}(23), 233102 June  (2014).

\bibitem{yasaitis_microwave_1975}
Yasaitis, J. and Rose, R.
\newblock {\em IEEE Transactions on Magnetics}{ \bf 11}(2), 434--436 March
  (1975).

\bibitem{singh_molybdenum-rhenium_2014}
Singh, V., Schneider, B.~H., Bosman, S.~J., Merkx, E. P.~J., and Steele, G.~A.
\newblock {\em Applied Physics Letters}{ \bf 105}(22), 222601 December  (2014).

\bibitem{kozinsky_tuning_2006}
Kozinsky, I., Postma, H. W.~C., Bargatin, I., and Roukes, M.~L.
\newblock {\em Applied Physics Letters}{ \bf 88}(25), 253101 June  (2006).

\bibitem{hoehne_damping_2010}
Hoehne, F., Pashkin, Y.~A., Astafiev, O., Faoro, L., Ioffe, L.~B., Nakamura,
  Y., and Tsai, J.~S.
\newblock {\em Physical Review B}{ \bf 81}(18), 184112 May  (2010).

\bibitem{venkatesan_dissipation_2010}
Venkatesan, A., Lulla, K.~J., Patton, M.~J., Armour, A.~D., Mellor, C.~J., and
  Owers-Bradley, J.~R.
\newblock {\em Physical Review B}{ \bf 81}(7), 073410 February  (2010).

\bibitem{imboden_dissipation_2014}
Imboden, M. and Mohanty, P.
\newblock {\em Physics Reports}{ \bf 534}(3), 89--146 January  (2014).

\bibitem{esquinazi_tunneling_1998}
Esquinazi, P., editor.
\newblock {\em Tunneling {Systems} in {Amorphous} and {Crystalline} {Solids}}.
\newblock Springer Berlin Heidelberg, Berlin, Heidelberg,  (1998).

\bibitem{esquinazi_acoustic_1986}
Esquinazi, P., Ritter, H.~M., Neckel, H., Weiss, G., and Hunklinger, S.
\newblock {\em Zeitschrift für Physik B Condensed Matter}{ \bf 64}(1), 81--93
  March  (1986).

\bibitem{singh_negative_2015}
Singh, V., Shevchuk, O., Blanter, Y.~M., and Steele, G.~A.
\newblock {\em arXiv:1508.04298 [cond-mat]}{ \bf } August  (2015).
\newblock arXiv: 1508.04298.

\end{thebibliography}
\end{document}


\title{Supplementary Material : Mechanical dissipation in MoRe superconducting metal drums}

\author{S. Yanai}
\author{V. Singh}
\author{M. Yuan}
\author{M. Gely}
\author{S.~J.,~Bosman}
\author{G.~A.,~Steele}
\affiliation{Department of Quantum Nanoscience,
Kavli Institute of Nanoscience, Delft University of Technology, Lorentzweg 1, 2628 CJ Delft, The Netherlands.}

\affiliation{Kavli Institute of NanoScience, Delft University of Technology, PO Box 5046, 2600 GA, Delft, The Netherlands.}

\date{\today}

\maketitle


\section{Fabrication details}
The fabrication process is mainly based on using PMGI as the sacrificial layer together with its compatibility with PMMA based resist. Steps of the fabrication process are illustrated in Fig.~1. 

\begin{figure}
\includegraphics[width=140mm]{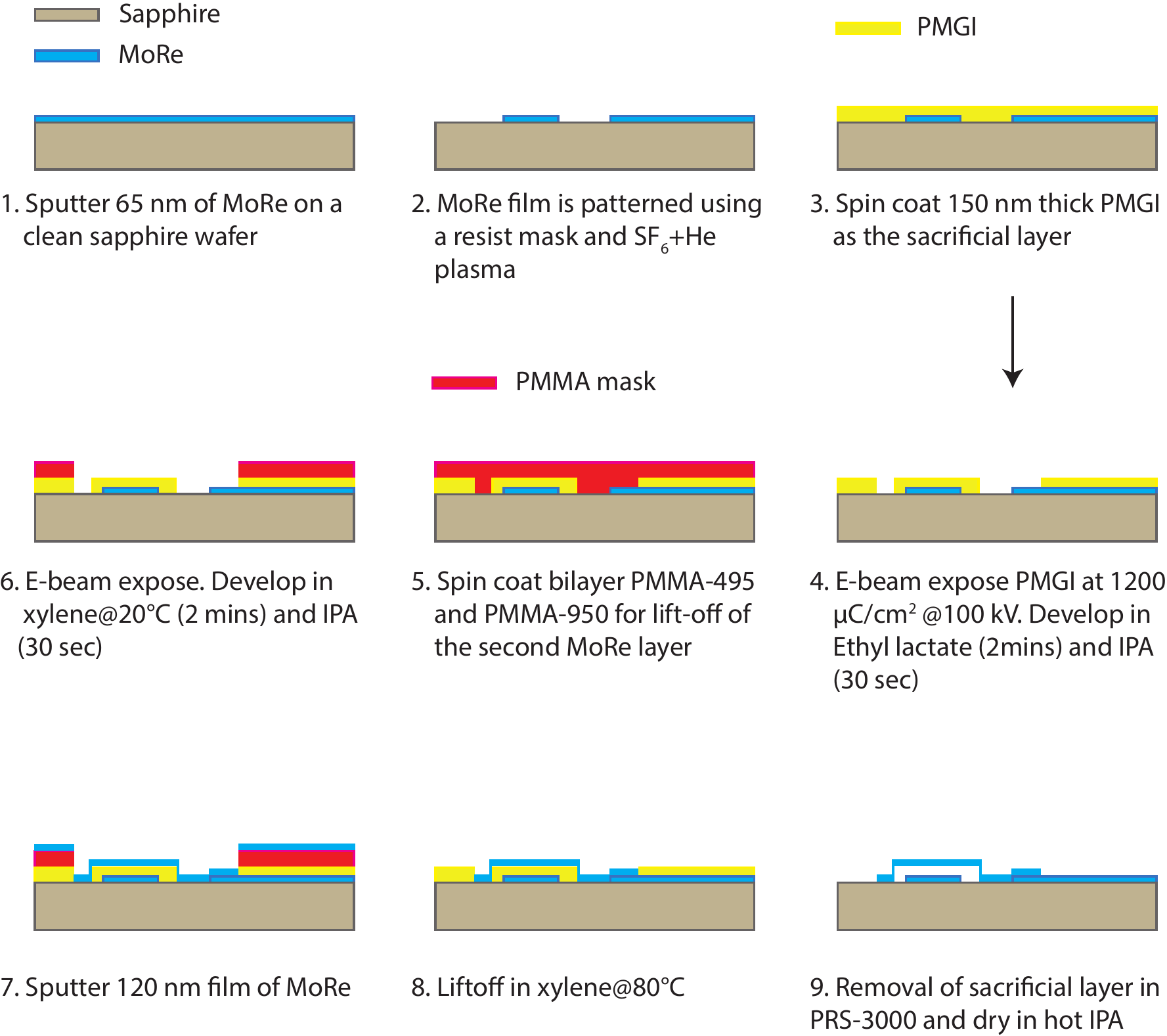}
\caption{An illustration of complete fabrication process with brief outline. }\label{fig0}
\end{figure}

\section{Characterization of the cavity}

For a single port cavity, the reflection coefficient $S_{11}$ is given by,

\begin{equation}
|S_{11}|(\omega)=\left|1-\frac{\kappa_e}{-i(\omega-\omega_c)+(\kappa_i+\kappa_e)/2}\right|
\end{equation}
where $\kappa_e$ is the external coupling rate, $\kappa_i$ is the internal dissipation rate, and $\omega_c$ is the cavity resonance frequency. Fig.~2(a) shows the measurement of $S_{11}$ at $T$~=~14~mK. The gray curve in Fig.~2(a) is the numerically fitted curve using equation 1. To differentiate between over-coupling ($\kappa_e > \kappa_i$) and under-coupling ($\kappa_i > \kappa_e$) limits, one must measure the phase $arg(S_{11}(\omega))$ of the reflection coefficient. Equivalently, one can also parametrically plot two quadratures $X = Re(S_{11})$ and $Y = Im(S_{11})$ against each other. For an overcoupled cavity the resonance circle appears to enclose the origin as shown in Fig~2(b). The additional circle (larger diameter) is due to the phase wrapping due the finite electrical length between the device and the circulator.

\begin{figure}
\includegraphics[width=120mm]{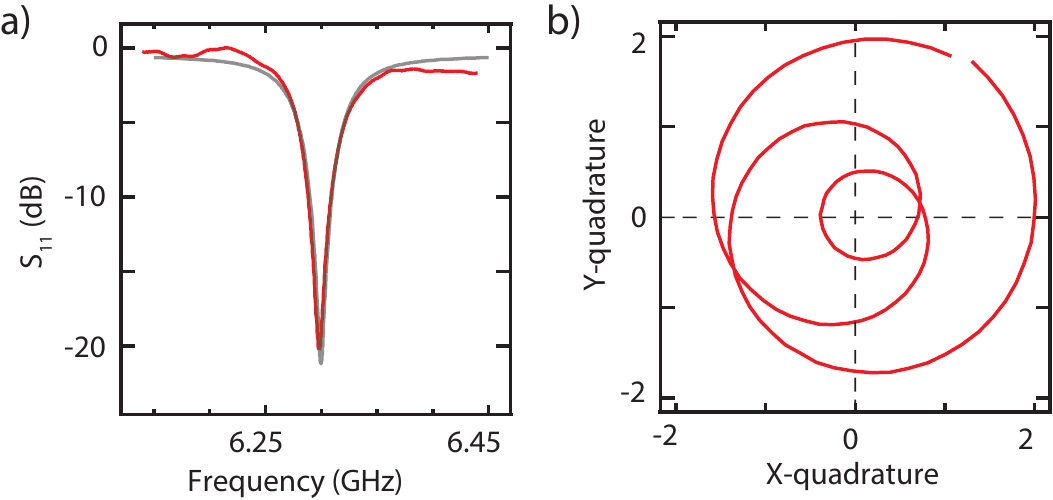}
\caption{(a) Measurement of the normalized reflection coefficient $S_{11}$ of the cavity at base temperature (red curve). The gray curve is the fit to the data yielding an internal dissipation rate $\kappa_i=2\pi\times25.8$~MHz and an external coupling rate $\kappa_e=2\pi\times31.0$~MHz. (b) Two quadratures of $S_{11}$ (raw data) plotted against each other, reflecting the over-coupled behavior of the cavity.  }\label{fig1}
\end{figure}

\section{Homodyne detection scheme }

To detect the mechanical motion, we use the superconducting microwave cavity in a homodyne detection setup. The mechanical resonator is actuated by capacitive forces by adding a radio-frequency signal $V_{AC}$ (near the mechanical resonance frequency) and a DC voltage $V_{DC}$ to the feedline (using a bias-tee at low temperature). Due to electrostatic force, the mechanical resonator experiences a modulating force of magnitude given by $\frac{dC_c}{dx}V_{DC}V_{AC}$, where $C_c$ is the capacitance between the mechanical resonator and the feedline electrode.

To detect the mechanical motion, we use the cavity as an interferometer. The cavity is driven by a signal at its resonance frequency $\omega_c$ acting as the carrier signal. The motion of the mechanical resonator (at frequency $\omega_m$) modulates this signal and produces mechanical sidebands at frequencies $\omega_c\pm\omega_m$. The sideband signal along with the carrier is amplified by a low noise amplifier and then passed through a high-pass filter (3~GHz-9~GHz) before entering the second amplifier at room temperature. The amplified signal is then mixed down to RF frequencies by using a mixer (MiniCircuits-ZMX8GH+). To avoid any phase drifts, we have used the same source to drive the local oscillator (LO) port of the mixer. The mixed-down RF signal is filtered and amplified further before entering the spectrum analyzer. The spectrum analyzer is used in tracking mode with the signal generator used to mechanically drive the resonator, thus allowing to record the response as function of actuation frequency. A complete schematic of the measurement scheme is shown in Fig~3.

\begin{figure}
\includegraphics[width=160mm]{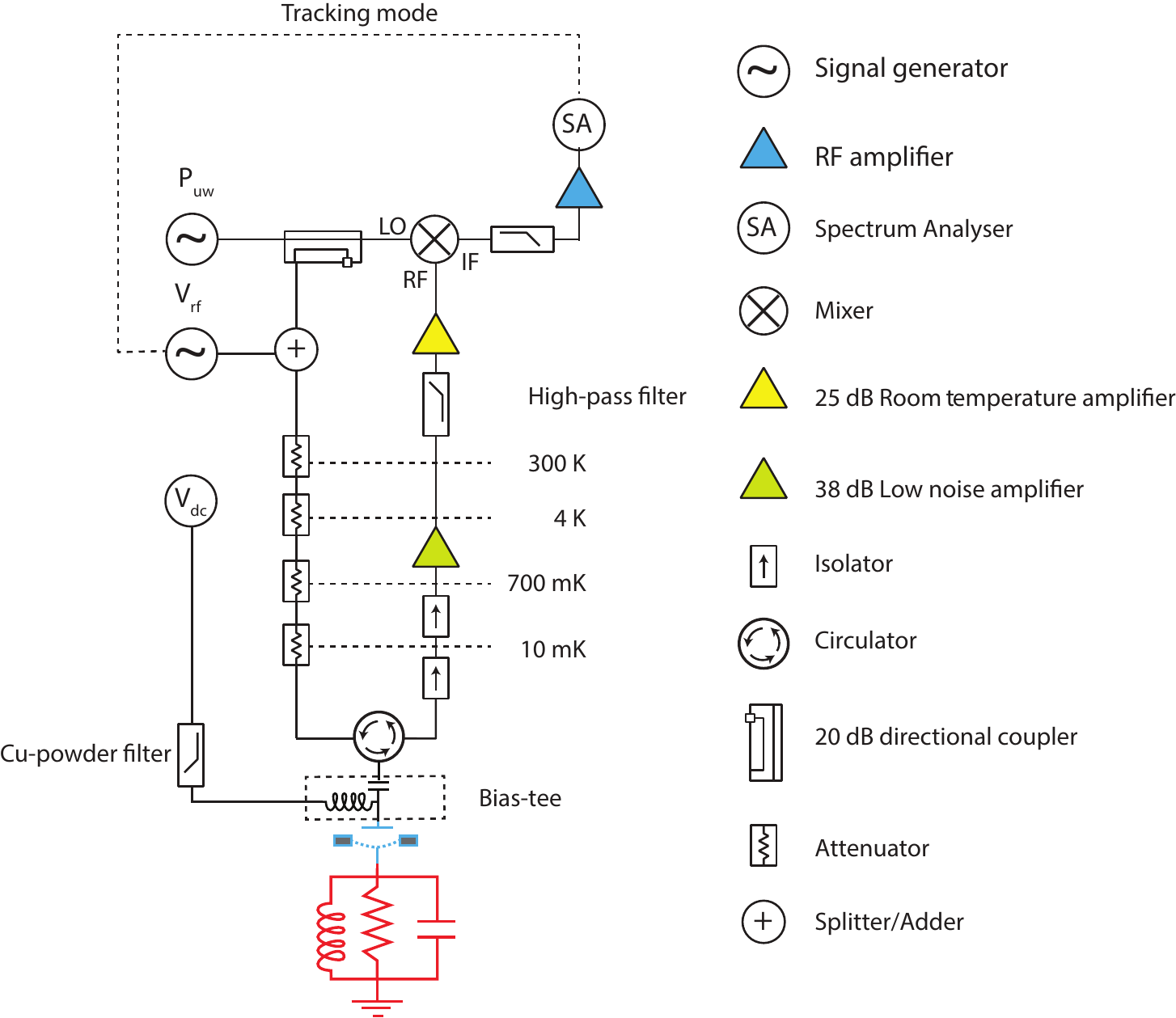}
\caption{Schematic diagram of the homodyne detection setup. }\label{fig2}
\end{figure}

\section{Fitting the asymmetric response with Fano-function}

Due to finite isolation of the circulator for low frequency RF signals (used for mechanical driving), part of the RF signal leaks directly into the output measurement chain, producing frequency independent constant background signal in the measurement. This background signal interferes with the signal generated due to the mechanical motion of the resonator, giving a Fano-lineshape in the measured data. Taking this background signal into account, the measured response can be fitted to the following Fano-form, also known as skewed-Lorentzian
\begin{equation}
\tilde{V}_{measured} = \frac{V_0}{1-2iQ_m\frac{f-f_m}{f_m}} + \alpha e^{i\phi}
\end{equation} 
where $f_m$ is the mechanical resonant frequency, $Q_m$ is the mechanical quality-factor, $\alpha e^{i\phi}$ is a small background signal with amplitude  $\alpha$ and phase $\phi$.

\section{Model for the estimation of the mechanical amplitude}

The combined effect of charging the transmission line and applying an AC-voltage leads to a force on the drum given by:
\begin{equation}\label{eq:dynamicsmech}
F = \frac{1}{2}\frac{\partial C}{\partial x}V^2 =
\frac{1}{2}\frac{\partial C}{\partial x}(V_{DC}+ V_{AC})^2 =
\frac{1}{2}\frac{\partial C}{\partial x}(V_{DC}^2+2V_{DC}V_{AC}+
V_{AC}^2)
\end{equation}
The first term $V_{DC}^2$ is a constant force which will attract the drum, increasing the stress of its membrane and thus lowering its frequency quadratically. The last term $V_{AC}^2$ can be neglected. We will call the second term the AC driving force $F_{AC}$. By assuming the capacitance of the drum to the transmission line to be that of a vacuum-gap, parrallel plate capacitor, we can write:
\begin{equation}
    F_{AC}= V_{AC}V_{DC}\frac{\partial}{\partial x}(\frac{\epsilon_0 A}{\bar{d}+x})
\end{equation}
Where $\epsilon_0$ is the permittivity of vacuum, $A$ is the area of the capacitive plates, $\bar{d}$ is the gap between the plates after having taken into account the constant displacement induced by the gate voltage. $x$ is the vertical displacement from this equilibrium position and is taken as a small quantity $x<<\bar{d}$ resulting in the expression for the AC drive:
\begin{equation}
    F_{AC}= -V_{AC}V_{DC}(\frac{\epsilon_0 A}{\bar{d}^2})
    \label{eq:AC_force}
\end{equation}
Assuming the drum behaves as a damped harmonic oscillator with an effective mass $m$, eigen frequency $\omega_m$ and quality factor $Q_m$, the amplitude of the mechanical oscillations can be related to the strength of the (resonant) drive:
\begin{equation}
    x_0 = \frac{F_{AC}Q_m}{m\omega_m^2}
    \label{eq:x_0}
\end{equation}
In this analysis the effective mass is taken to be the total mass of the drum (this choice defines the amplitude $x_0$ which is otherwise ambiguous due to the geometry of the drum). We calculate the amplitude of the AC-voltage from the input power at the device $P_{DUT}$
\begin{equation}
    V_{AC} = \sqrt{2Z_0P_{DUT}}
    \label{eq:AC_voltage}
\end{equation}
where $Z_0 = 50~\Omega$ is the characteristic impedance of our line. This power can be calculated from the output power of our signal generator and an estimation of the attenuation on the input measurement chain. The following steps are then taken:
\begin{itemize} 
\item We convert the input power at our signal generator to the force acting upon the drum using \ref{eq:AC_voltage} and \ref{eq:AC_force}
\item We associate the peak measured power $p_0$ with the amplitude mechanical oscillation $x_0$ using \ref{eq:x_0}
\item We convert the power received off-resonance $p$ to a mechanical amplitude $x = x_0 \sqrt{p /p_0}$
\end{itemize}

Alternatively, one could also use the measured homodyne signal to work out the amplitude of the mechanical resonator under a coherent drive. For a single port cavity driven at its resonant frequency ($\Delta$ = 0), amplitude of the drumhead resonator can be written as, $x_p^2={\frac{\omega_m^2+\kappa^2/4}{(\eta G)^2}}\frac{P_o}{P_i}$, where $G$ is the cavity pull-in parameter, $\eta=\kappa_e/\kappa$ is the coupling fraction, $P_o$ is the homodyne power (after removing gain, loss of the output chain, and conversion loss of the mixer), and $P_i$ is the injected power outside the cavity.

\section{Additional data from second device}

Additional data from second device showing the nonlinear damping of the mechanical resonator is shown in Fig.~4.
\begin{figure}
\includegraphics[width=160mm]{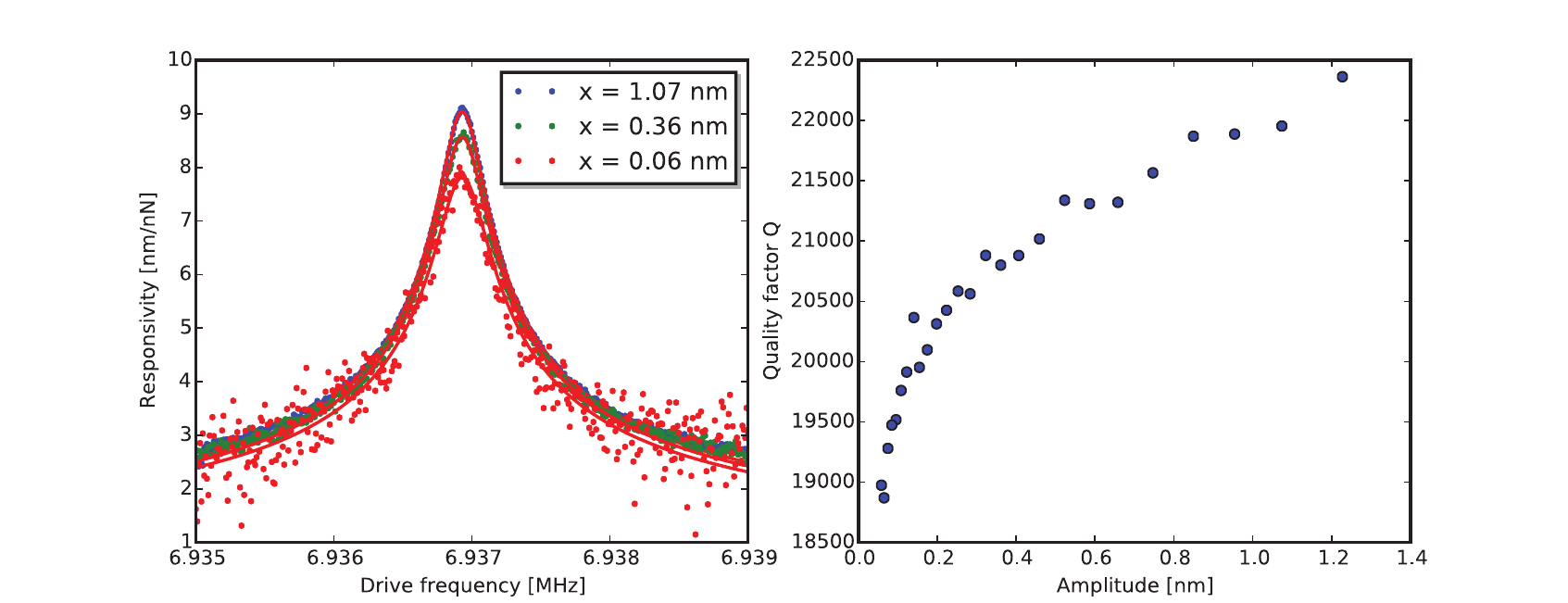}
\caption{Mechanical response and mechanical quality factor as a function of power}\label{fig4}
\end{figure}

\section{Estimation of the gap between the drum and the bottom electrode}

The gap between the drum and the bottom electrode can be estimated from the total inductance of the high impedance resonator. The total inductance of the microstrip resonator is estimated to be 3.43~nH using finite element modeling and considering kinetic inductance of Mo-Re. Total capacitance is calculated from geometric inductance and the cavity frequency.
Coupling quality factor in single port cavity is given as,
\begin{equation}
{Q}_{ext} = \frac{C_l+C_{c}}{\omega_{c} C_{c}^2 Z_{0}}
\end{equation} 
where $C_{c}$ is coupling capacitance, $\omega_{c}$ is resonance frequency  of the cavity, $Z_{0}$ is input impedance of transmission line, and $C_l$ is lumped capacitance of the resonator line.
From the parameters of cavity characterization, coupling capacitance is 21.5~fF. Assuming a parallel plate capacitor model, we found the mean position of the drum head resonator  $\bar{d}$ to be 291~nm for the device represented in Fig.~4 of main text. \\

Alternatively, one can perform opto-mechanically induced absorbtion (OMIA) \cite{ Vibhor_graphene_optomechanics_2014} leading to a measurement of the cooperativity $C = 4g^2/\kappa\Gamma$ where $\kappa$ is the cavity line-width, $\Gamma$ the mechanical line-width and $g=G \sqrt{n_d} \sqrt{\hbar/2m\omega_m}$ is the opto-mechanical coupling strength where the average number of photons in the cavity $n_d$ in that particular measurement is calculated from our estimation of the input attenuation. OMIA thus provides a measure of the cavity pull-in parameter $G=d\omega_c/dx = $8.53$\times$10$^{14}$~Hz/m. For a $\lambda /4$ resonator, the pull-in parameter assuming a parrallel plate capacitor is given by:
  \begin{equation}
    G = \frac{2 Z_0 \omega_r^2}{\pi}\frac{\epsilon A}{\bar{d}^2}
  \end{equation}
from this we estimate the mean position of the drum head resonator $\bar{d}$ to 481~nm for the device represented in Fig.~4 of the supplementary materials.

\section{Parameters for the devices shown in the main text (device 1) and in supplemental materials (device 2)}

\begin{table}
\caption{Important parameters of the devices}
\begin{center}
\begin{tabular}{ | m{8cm} | m{3cm}| m{3cm}| } 
\hline
\textbf{Parameters} & \textbf{Device 1} & \textbf{Device 2} \\
\hline
Density of MoRe& 14.5 g/cm$^3$& 14.5 g/cm$^3$  \\ 
\hline
Diameter of drum & 30~$\mu$m & 30~$\mu$m \\ 
\hline
Gap between the drumhead and the feedline & 291~nm & 481~nm \\ 
\hline
Thickness of the mechanical resonator & 125~nm & 131~nm \\ 
\hline
Equivalent lumped inductance of the cavity & 3.43 nH & 2.37 nH \\ 
\hline
Drumhead capacitance & ~21.5~fF & 12.6~fF \\ 
\hline
Cavity pull-in parameter $\frac{G}{2\pi}$ & 1.2$\times$10$^{15}$~Hz/m & 8.53$\times$10$^{14}$~Hz/m \\ 
\hline
Total attenuation on low frequency drive line & 65 dB & 55 dB \\ 
\hline

\end{tabular}
\end{center}

\end{table}